\documentclass{ptephy_v1}

\preprintnumber{XXXX-XXXX} 

\usepackage{arydshln}
\usepackage{color}
\usepackage{multirow}
\usepackage{lscape}

\begin{document}

\title{Development of a method for  measuring rare earth elements in the environment for future experiments with gadolinium loaded detectors}


\author[1,*]{S.~Ito}
\affil{Okayama University, Faculty of Science, Okayama 700-8530, Japan \email{s-ito@okayama-u.ac.jp}}

\author[2]{T.~Okada}
\affil{Kamioka Observatory, Institute for Cosmic Ray Research, University of Tokyo, Kamioka, Gifu 506-1205}

\author[3]{Y.~Takaku} 
\affil{Institute for Environmental Sciences, Department of Radioecology, Aomori, 039-3212, Japan}

\author[1]{M.~Harada}

\author[2,4]{M.~Ikeda}
\affil[4]{Kavli Institute for the Physics and Mathematics of the Universe (WPI), the University of Tokyo, Kashiwa, Chiba, 277-8582, Japan}

\author[2,4,5]{Y.~Kishimoto}
\affil[5]{Present address: Research Center for Neutrino Science, Tohoku University, Sendai 980-8578, Japan}

\author[1]{Y.~Koshio}

\author[2,4]{M.~Nakahata}

\author[2,4]{Y.~Nakajima}

\author[2,4]{H.~Sekiya}


\begin{abstract}%
Demand to use gadolinium (Gd) in detectors is increasing in the field of elementary particle physics, especially neutrino measurements and dark matter searches. 
Large amounts of Gd are used in these experiments. 
Therefore, to access the impacts of Gd onto the environments, it is becoming important to measure the baseline concentrations of Gd in the environments. 
The measurement of the baseline concentrations, however, is not easy due to interferences by other elements. 
In this paper, a method for measuring the concentrations of rare earth elements including Gd is proposed. 
In the method, an inductively coupled plasma-mass spectrometry is utilized after collecting the dissolved elements in chelating resin. 
Results of the ability to detect anomalous concentrations of rare earth elements in river water samples in the Kamioka and Toyama areas are also reported.    

\end{abstract}

\subjectindex{xxxx, xxx}

\maketitle

\section{Introduction}

\subsection{Gadolinium loaded detectors}

The rare earth element gadolinium (Gd) has one of the largest thermal neutron capture cross sections. 
Once a neutron is captured by Gd, it emits multiple $\gamma$-rays with total energy of about 8 MeV (Ref. \cite{Beacom}). 
These $\gamma$-rays can be used as signals to detect a capture event; therefore, Gd is often included in neutrino detectors. 
For example, the RENO experiment (Ref. \cite{RENO1}), the Double Chooz experiment (Ref. \cite{DC}), and the Daya Bay experiment (Ref. \cite{DB}), measuring the neutrino mixing matrix parameter ${\theta}_{13}$, use Gd-loaded liquid scintillators to detect neutrons from inverse beta decay of reactor neutrinos (${\overline{\nu}}_{e}+p{\rightarrow}e^++n$). 

Many projects using Gd-loaded water Cherenkov detectors are currently in development. 
For example, the Super-Kamiokande Gd (SK-Gd) project aims to detect supernova relic neutrinos by loading Gd into the Super-Kamiokande (SK, Ref. \cite{NIMA}) water Cherenkov detector (Ref. \cite{Beacom}).  
Additionally, WATCHMAN (Ref. \cite{WATCHMAN}) and ANNIE (Ref. \cite{ANNIE}) use Gd-loaded water Cherenkov detectors. 
Among dark matter experiments, the XENON-1T detector will be upgraded to the XENON-nT detector with a neutron veto system based on a Gd-loaded water Cherenkov detector (Ref. \cite{Xenon}). 
As applications of Gd are getting wider especially in physics researches, it is becoming more important to accurately measure Gd in the environment from the viewpoints of keeping the concentration of Gd constant in the detector and of preventing Gd leakage into the environment.

\subsection{Surveys of anomalies of rare earth elements using ICP-MS}

The concentrations of rare earth elements including Gd depend highly on suspended substances in natural waters. 
Additionally, the concentrations of rare earth elements generally show zigzag patterns that follow the Odd-Harkins rule (Ref. \cite{REEs}). 
Therefore, to detect possible anomalies of rare earth elements in the environment, the concentrations of all rare earth elements should be measured. 
To minimize the effects of suspended substances and the zigzag concentrations of rare earth elements, the concentration of each rare earth element is often normalized to the concentration of chondrites (stony meteorites) (Ref. \cite{Chondrite}). 
The normalized concentrations of rare earth elements tend to be distributed as smooth curve and deviations  from this smooth curve will indicate anomalies of rare earth elements. 

Inductively coupled plasma-mass spectrometry (ICP-MS)  is used widely to detect small amounts of elements in solutions thanks to its sensitivity. 
The general sensitivity of ICP-MS to rare earth elements\footnote[1]{An element of promethium (Pm) is also classified as rare earth elements, but could not be determined by ICP-MS since all of its isotopes are radioactive with short lifetime.} is $<0.1$ pg mL$^{-1}$. 
The concentrations of rare earth elements in the natural waters are expected to be 1$-$10 pg mL$^{-1}$, so ICP-MS is well suited for detecting anomalies.  
However, alkali metals (e.g. cesium (Cs)) and alkaline earth metals (e.g. barium (Ba)) are also dissolved in  natural water at higher concentrations than rare earth elements. 
Those elements can form polyatomic ions, such as barium oxide (BaO), the concentrations of rare earth elements will tend to be overestimated when their masses are similar. 
This interference is generally referred to as spectral interference. 
To minimize spectral interference and determine the concentrations of rare earth elements precisely, we developed a  chemical separation and preconcentration procedure.  
As a test case, rare earth elements in the rivers of the Kamioka and Toyama areas of Japan were measured using the developed method. 
Details of the measurement methods and results are reported in the following sections.

\section{Instrument, experimental equipment, and reagents}

The Agilent 7900 was used to detect rare earth elements (Ref. \cite{Agilent}). 
This ICP-MS was housed in a clean room in the Kamioka mine. 

\begin{figure}[htbp]
\centerline{\includegraphics[width=10cm, trim= 8cm 6cm 2.5cm 5cm, clip]{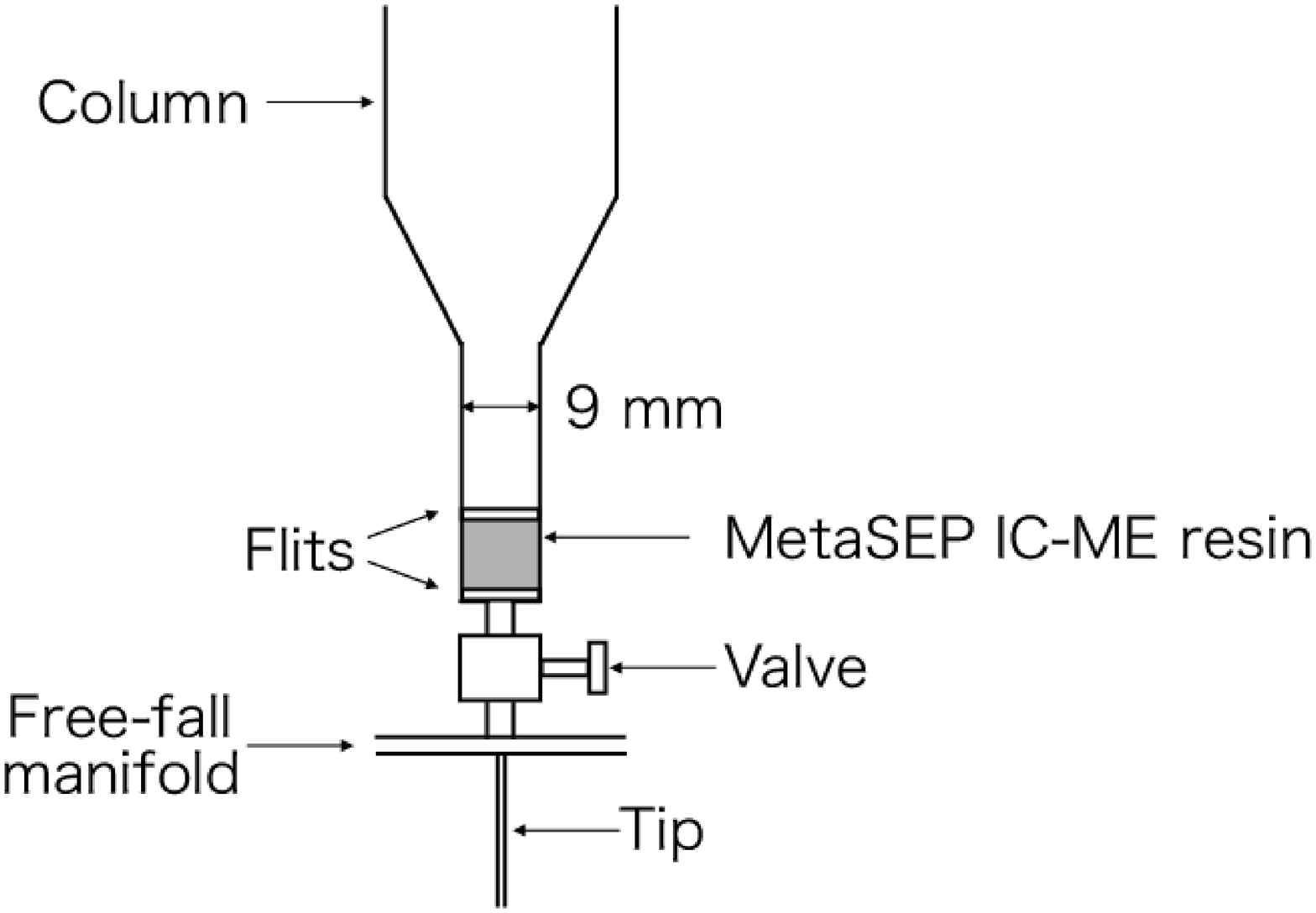}}
\caption{Schematic of the experimental setup.}
\label{fig:column}
\end{figure}

To minimize spectral interference from alkali metals and alkaline earth metals and enhance the sensitivity of the ICP-MS measurements, chemical separation and preconcentration with chelating resin is often used (e.g. Ref. \cite{Resin1, Resin2, Resin3}). 
The MetaSEP IC-ME chelating resin was used in the experiments (GL Sciences Inc., Ref. \cite{GLS}). 
At pH$\simeq$5, rare earth elements are strongly adsorbed by the resin while alkali metals and alkaline earth metals are not. 
Figure \ref{fig:column} shows a schematic of the experimental setup. 
A column with a volume of about 20 mL and an inside diameter of 9 mm (LSC-$\phi$9, GL Sciences Inc., Ref. \cite{GLS}) was used. 
A valve was added at the bottom end of the column to adjust the flow rate. 
This setup was connected to a tip, and fixed to a free-fall manifold. 
The resin was sandwiched between two flits to fix it. 

To produce solutions with low contamination, ultra-pure SK water (Ref. \cite{NIMA}) and ultra-high-purity analytical grade 68\% nitric acid (HNO$_3$, TAMAPURE AA-100 from Tama Chemicals, Ref. \cite{TAMA}) were used. 
The reagents for adjusting the pH of the sample solutions were ultra-high-purity analytical grade 30\% acetic acid (CH$_3$COOH, TAMAPURE AA-100, Tama Chemicals, Ref. \cite{TAMA}) and 20\% ammonia solution (NH$_3$, TAMAPURE AA-100, Tama Chemicals, Ref. \cite{TAMA}). 
The ICP-MS was calibrated and the recovery rate of rare earth elements was estimated using a standard solution of 16 high-purity rare earth elements (XSTC-1, SPEX Inc., Ref. \cite{Spex}). 
A high-purity 29-elements standard solution (XSTC-331, SPEX Inc., Ref. \cite{Spex}) was also used to test the recovery. 
Electronic (EL) grade 70\% HNO$_3$ (Wako Pure Chemical Industries Ltd., Ref. \cite{Wako}) was used to prewash the experimental equipment. 

To minimize the undissolved particles in the river waters, a membrane filter with pore size of 0.45 $\mu$m (Omnipore filter, Merck Millipore Inc., Ref. \cite{Merck}) was used. 
All the bottles used for collection of river water, production of the solutions, preservation of the samples, and ICP-MS measurements were made of polypropylene. 

\section{Experimental method}

\subsection{Sampling and pretreatment}

\begin{figure}[htbp]
\centerline{\includegraphics[width=12cm, trim= 7cm 0cm 7cm 0cm, clip]{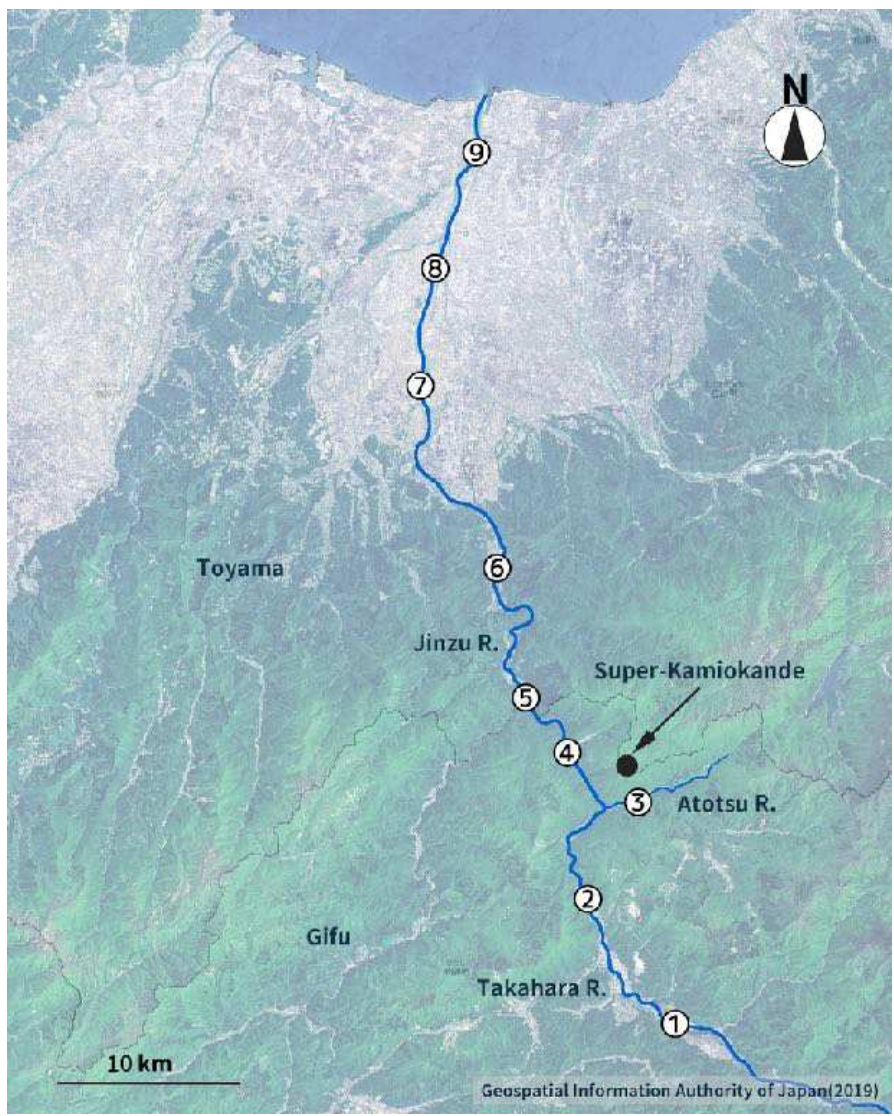}}
\caption{Map of the sampling points (Ref. \cite{Geo}). For the sake of visibility, the size of the river is  illustrated as larger than its actual size.}
\label{fig:MapRiver}
\end{figure}

On November 15th and 16th, 2018, river samples were taken from upstream in the Takahara river to  downstream in the Jinzu river, and from the Atotsu river in which SK area's water flows. 
The sampling points are marked in Figure \ref{fig:MapRiver}. 

River water was collected in bottles after they were washed with 1 mol L$^{-1}$ HNO$_3$ (EL grade) solution and rinsed with ultra-pure SK water. 
The collected samples were immediately filtered using membrane filters and acidified to pH$\simeq$1 by HNO$_3$, and were stored in the clean room until testing.

\subsection{Procedure of the chemical separation and preconcentration}

The resin and equipment were likely contaminated during the production process. 
Thus, all the equipment was soaked in 1 mol L$^{-1}$ HNO$_3$ (EL grade) solution for at least one night and rinsed using the ultra-pure SK water before use. 

0.3 g of the resin was loaded into the column and was set as illustrated in Figure \ref{fig:column}. 
To wash the resin, inside of the column, the valve, and the tip, 20 mL of 3 mol L$^{-1}$ HNO$_3$ (EL grade) was loaded into the column and extracted through the tip. 
This procedure was repeated three times. 
Then, 20 mL of 3 mol L$^{-1}$ HNO$_3$ was loaded into the column and passed through the resin for a reference sample. 

The pH of the sampled river water was adjusted to $\sim$5 using CH$_3$COOH and NH$_3$ solutions. 
The resin was conditioned using 20 mL of the conditioning solution with pH$\simeq$5, which was made of the ultra-pure SK water, HNO$_3$, CH$_3$COOH, and NH$_3$ solutions. 
Then, 50 mL of the sample solution was loaded into the column. 
The flow rate was adjusted to about 0.5 mL min$^{-1}$ using the valve. 
With this procedure, most of the rare earth elements are adsorbed by the resin while the alkali metals and alkaline earth metals pass through the resin. 

To rinse any remaining alkali metals and alkaline earth metals out of the resin, 20 mL of the conditioning solution was loaded into the column. 
The adsorbed rare earth elements were eluted using 10 mL of 3 mol L$^{-1}$ HNO$_3$ solution. 
All rare earth elements are concentrated by approximately a factor of five through this eluting procedure. 
The eluted solution was collected and measured with the ICP-MS. 
The total time for the chemical separation and preconcentration procedure, and the ICP-MS measurement was about one day.

\section{Results and discussions}

\begin{figure}[htbp]
\centerline{\includegraphics[width=11.5cm]{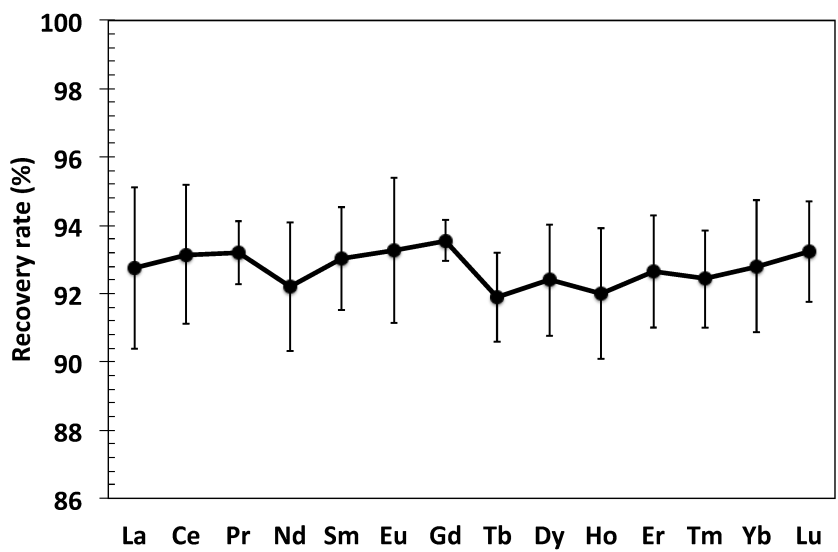}}
\caption{Evaluated recovery rates of all rare earth elements measured using XSTC-1 solution.}
\label{fig:Recovery}
\end{figure}

Before the measurements of the river water samples, the recovery rates of rare earth elements were evaluated using 10 pg mL$^{-1}$ XSTC-1 and 10 ng mL$^{-1}$  XSTC-331 solutions. 
Figure \ref{fig:Recovery} shows the recovery rates of all rare earth elements. 
All were recovered $>90$\%, while alkali metals and alkaline earth metals like Cs and Ba were eliminated to  $<0.1$\%. 

\begin{figure}[htbp]
\centerline{\includegraphics[width=11.5cm]{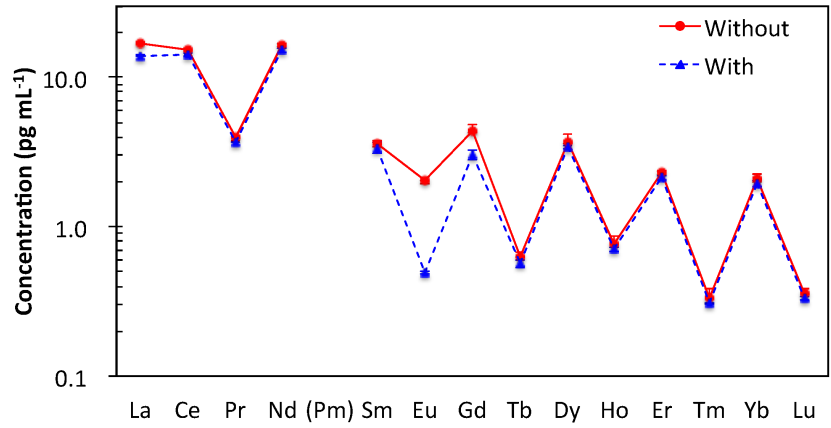}}
\caption{Comparison of sample 1 between the measurements with (red circles with solid red line) and without (blue triangles with dashed blue line) the developed chemical separation and preconcentration procedure. The errors are enough small and the illustrated error bars are too small to be visible.}
\label{fig:ConcComparison}
\end{figure}

Figure \ref{fig:ConcComparison} shows a comparison of results from sample 1 when it was prepared with and without the chemical separation and preconcentration procedure. 
Typical spectral interference is shown in europium (Eu). 
The concentration of Eu as measured with $^{153}$Eu without any chemical separation or preconcentration procedure was 2.0 pg mL$^{-1}$, while the concentration determined with the developed method was 0.5 pg mL$^{-1}$. 
This gap is explained by spectral interference from $^{137}$Ba$^{16}$O. 
The concentrations of lanthanum (La) and Gd in the direct measurements were 16.9$\pm$0.2 pg mL$^{-1}$ and 4.4$\pm$0.5 pg mL$^{-1}$, respectively. 
These values are also higher than the results found with the developed chemical separation and preconcentration procedure (14.0$\pm$0.3 pg mL$^{-1}$ for La and 3.0$\pm$0.2 pg mL$^{-1}$ for Gd, respectively). 
These gaps also imply the possibility for spectral interferences in La and Gd measurements. 
If the sample solutions are directly measured by the ICP-MS without any chemical extraction and preconcentration procedure, the elements may be overestimated due to significant spectral interference. 
On the other hand, the spectral interference to rare earth elements can be efficiently minimized using the developed method.

\begin{figure}[htbp]
\centerline{\includegraphics[width=15.5cm, trim= 0cm 4.8cm 0cm 5cm, clip]{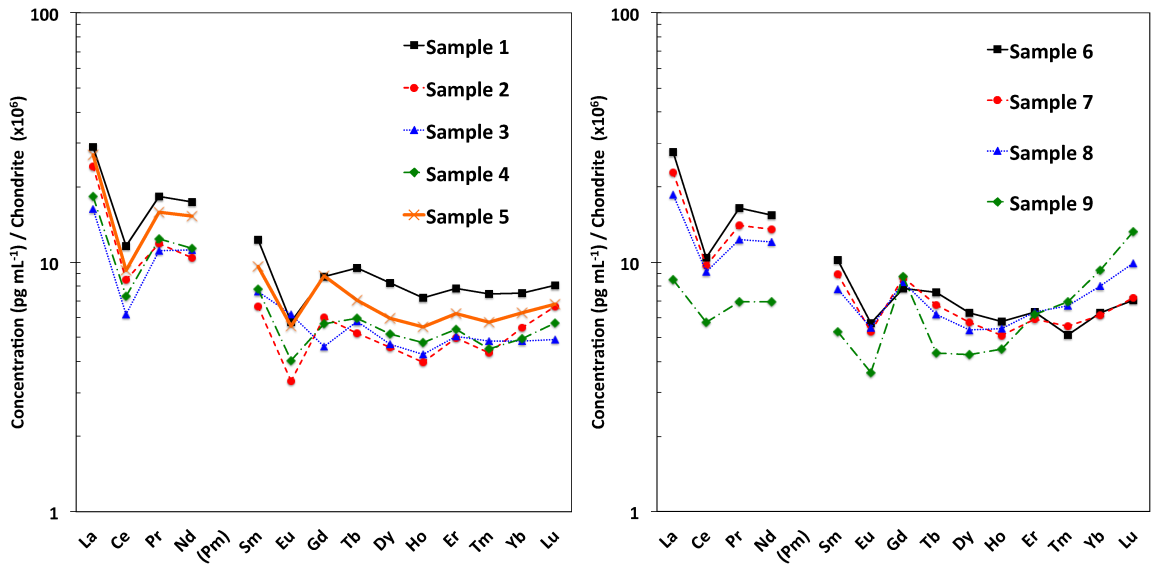}}
\caption{Normalized concentrations of rare earth elements in samples 1 to 5 (left) and samples 6 to 9 (right).}
\label{fig:Normalized_all}
\end{figure}

Table \ref{tab:results} lists the results of the measurements of rare earth elements in the river waters using the developed method. 
For the concentrations of rare earth elements, the values of the blanks were subtracted and the recovery rate was corrected. 
Figure \ref{fig:Normalized_all} shows the concentrations of rare earth elements in all samples normalized to  chondrite concentration. 
The normalized concentrations should be distributed on a smooth curve since the natural zigzag pattern of rare earth elements is minimized by the normalization. 
However, negative anomalies of cerium (Ce) and Eu can also be seen in all samples expect for sample 3. 
Those anomalies are caused by different oxidation rates. 
The valence of all rare earth elements is usually +3, however, Ce and Eu can have two different valences, +3 and +4 for Ce and +2 and +3 for Eu, under different redox conditions. 
Ions of Ce$^{4+}$ and Eu$^{2+}$ are more easily oxidized, resulting the negative anomalies.

A positive anomaly of Gd can be seen in sample 9. 
Because compounds of Gd are used widely as a contrast medium for magnetic resonance imaging in medical diagnostics, it is often observed in urban regions (Ref. \cite{Pollution1, Pollution2}). 

All rare earth elements can be detected using the developed method, and no anomalous concentrations of rare earth elements were found near SK. 
After normalization, the concentration of each rare earth element can be compared and the excess of rare earth elements can be  examined. 
Therefore, this method is appropriate for assessing the concentration of Gd in the environment. 

\begin{landscape}
\begin{table}[htbp]
\begin{center}
\caption{Measurements of the river samples (unit: pg mL$^{-1}$). Values without error had the negligible error.}
\begin{tabular}{|c|ccccccccc|}\hline
\multirow{2}{*}{Elements} & \multicolumn{9}{c|}{Sample} \\ \cline{2-10}
 & 1 & 2 & 3 & 4 & 5 & 6 & 7 & 8 & 9 \\ \hline
La & 14.0$\pm$0.3 & 11.7$\pm$0.5 & 7.9$\pm$0.2 & 8.9$\pm$0.5 & 13.1$\pm$0.3 & 13.4$\pm$0.5 & 11.1$\pm$0.3 & 9.0$\pm$0.5 & 4.1$\pm$0.2 \\
Ce & 14.3$\pm$0.4 & 10.5$\pm$0.5 & 7.6$\pm$0.2 & 9.0$\pm$0.4 & 11.4$\pm$0.3 & 12.8$\pm$0.6  &  11.9$\pm$0.4 & 11.3$\pm$0.5 & 7.0$\pm$0.3 \\
Pr & 3.7 & 2.4 & 2.2$\pm$0.1 & 2.5$\pm$0.1 & 3.2 & 3.3$\pm$0.1 & 2.8 & 2.5$\pm$0.1 & 1.4$\pm$0.1  \\
Nd & 15.3$\pm$0.6 & 9.2$\pm$0.4 & 9.8$\pm$0.2 & 10.0$\pm$0.8 & 13.5$\pm$0.3 & 13.5$\pm$1.0 & 11.9$\pm$0.3 & 10.6$\pm$0.4 & 6.1$\pm$0.1 \\
Sm & 3.4$\pm$0.3 & 1.8$\pm$0.1 & 2.1$\pm$0.1 & 2.1$\pm$0.1 & 2.6$\pm$0.1 & 2.8$\pm$0.2 & 2.4$\pm$0.1 & 2.1 & 1.4$\pm$0.1  \\
Eu & 0.5 & 0.3 & 0.5 & 0.3 & 0.5 & 0.5 & 0.5 & 0.5 & 0.3  \\
Gd & 3.0$\pm$0.2 & 2.1$\pm$0.2 & 1.6$\pm$0.2 & 2.0$\pm$0.3 & 3.0$\pm$0.2 & 2.7$\pm$0.2 & 3.0$\pm$0.3 & 2.9$\pm$0.2 & 3.0$\pm$0.2  \\
Tb & 0.6 & 0.3 & 0.3 & 0.4 & 0.4 & 0.5 & 0.4 & 0.4 & 0.3  \\
Dy & 3.4$\pm$0.1 & 1.9$\pm$0.1 & 1.9$\pm$0.1 & 2.1$\pm$0.1 & 2.5$\pm$0.1 & 2.6$\pm$0.1 & 2.4$\pm$0.1 & 2.2$\pm$0.2 & 1.8$\pm$0.1  \\
Ho & 0.7 & 0.4 & 0.4 & 0.5 & 0.5 & 0.6 & 0.5 & 0.5 & 0.4  \\
Er & 2.1$\pm$0.1 & 1.4 & 1.4$\pm$0.1 & 1.5$\pm$0.1 & 1.7 & 1.7$\pm$0.2 & 1.6$\pm$0.1 & 1.7$\pm$0.1 & 1.7$\pm$0.1 \\
Tm & 0.3 & 0.2 & 0.2 & 0.2 & 0.2 & 0.2 & 0.2 & 0.3 & 0.3  \\
Yb & 1.9$\pm$0.1 & 1.4$\pm$0.1 & 1.3$\pm$0.1 & 1.3$\pm$0.1 & 1.6$\pm$0.1 & 1.6$\pm$0.2 & 1.6$\pm$0.1 & 2.1$\pm$0.2 & 2.4$\pm$0.1  \\
Lu & 0.3 & 0.3 & 0.2 & 0.2 & 0.3 & 0.3 & 0.3 & 0.4 & 0.5  \\ \hline
\end{tabular}
\label{tab:results}
\end{center}
\end{table}
\end{landscape}

\section{Conclusion}

New experiments that use Cherenkov detectors with Gd-loaded water are in development. 
A method of the assessment of Gd in the environment has been established with the chemical separation and preconcentration procedure described above. 
This procedure eliminates the spectrum interference in the ICP-MS measurements from alkali metals and alkaline earth metals, while retaining more than 90\% of all the rare earth elements. 
Consequently, all rare earth elements including Gd can be determined with precision of 0.1 pg mL$^{-1}$. 
This concentration corresponds to about 1 g day$^{-1}$ of Gd leakage from the SK tank to the environment. 

\ack
This work was supported by the JSPS KAKENHI Grants Grant-in-Aid for Scientific Research on Innovative Areas No. 26104006, Grant-in-Aid for Specially Promoted Research No. 26000003, Grant-in-Aid for Young Scientists No. 17K14290, and Grant-in-Aid for JSPS Research Fellow No. 18J00049.

\end{document}